\documentclass[aps,twocolumn,10pt,showpacs,showkeys]{revtex4}
\usepackage{amssymb}
\usepackage{bm}
\usepackage{graphicx}
\newcommand{\be}{\begin{eqnarray}}
\newcommand{\ee}{\end{eqnarray}}

\begin{document}
\draft
\title{Shell evolution of N=20 nuclei and Gamow-Teller strengths of $^{30,32,34}$Mg by the deformed QRPA}

\author{Eunja Ha \footnote{ejha@ssu.ac.kr} and Myung-Ki Cheoun  \footnote{Corresponding author : cheoun@ssu.ac.kr}}
\address{ Department of Physics,
Soongsil University, Seoul 156-743, Korea }

\begin{abstract}
Gamow-Teller (GT) strength distributions of Mg isotopes are investigated within a framework of the deformed quasi-particle random phase approximation(DQRPA).
We found that the N=20 shell closure in $^{28 \sim 34}$Mg was broken by the prolate shape deformation originating from the {\it fp}-intruder states. The shell closure breaking gives rise to a shift of low-lying GT excited states into high-lying states. Discussions regarding the shell evolution trend of single particle states around N=20 nuclei are also presented with the comparison to other approaches.
\end{abstract}

\pacs{\textbf{23.40.Hc, 21.60.Jz, 26.50.+x} }
\date{\today}

\maketitle
The structure of nuclei along the valley of stability is well understood in terms of the conventional magic numbers in the nuclear shell model.
In a region far from the stability, experimental studies have shown the breaking of the N=20 magic number in N=20 nuclei, such as $^{31}$Na, $^{32}$Mg, and $^{33}$Al nuclei located in so called the island of inversion \cite{Thib, Klotz, Motoba}. The breaking comes from the burrowing of $f_{7/2}$ and $p_{3/2}$ states below $d_{3/2}$ state \cite{Warb90}, which may originate from the monopole component of the nucleon-nucleon interaction by the tensor force \cite{Otsuka01}. The breaking of magic numbers in the inversion island is thought as one of the interesting properties of the exotic nuclei near drip line.

The neutron-rich nuclei around N=20 are thought to be strongly deformed because of the reduction of the N=20 shell gap.
The deformation in nuclei becomes more important than any other periods with the recent development of rare isotope accelerator facilities, from which one may perform lots of challenging experiments related to the exotic nuclei. Although they decay fractions of a second, their existence is imprinted on the nuclear abundances of stars through the successive nuclear reactions in the cosmos at the explosive stage of stellar evolution, such as supernovae (SNe) explosion  \cite{Haya04}.

For example, the nuclear reactions induced by the neutrinos emitted from the SNe are treated as important input data for the neutrino-process \cite{Heg,Ch10-2,Ch12}. Since the neutrinos emitted from a proto-neutron star may have tens of MeV energy high enough to excite the deformed nuclei, one needs to understand more precisely the high-lying (HL) excited states beyond one nucleon threshold. Among them, the GT states are of great importance because most of the charge exchange reactions are dominated by the GT transition. These HL GT excited states are in close association with the nuclear structure.

One of typical instances is the GT quenching problem. It says that the difference of total running sums $S_{\pm}$ for GT ($\pm$) transitions, $(S_- - S_+)$, is usually quenched compared to the Ikeda sum rule (ISR), $(S_- - S_+) = 3 ( N - Z)$. But, recent experimental data on HL GT states deduced by more energetic projectiles shed a new light on the GT states located above one nucleon threshold, whose contributions enable us to explain the quenching problem through the multi-particle and multi-hole configuration mixing \cite{Waka06}.

In order to describe neutron-rich nuclei and their relevant nuclear reactions occurred in the nuclear processes, one needs to
develop theoretical frameworks including explicitly the deformation
\cite{simkovic,saleh}. Ref. \cite{simkovic} used an effective separable force to the deformed quasi-particle random phase
approximation (DQRPA). Realistic two-body interaction was firstly exploited at Ref. \cite{saleh}.

In this paper, we apply our previous DQRPA \cite{Ha1,Ha2} to the nuclei in the N=20 inversion island. Starting from the deformed Wood Saxon potential \cite{Hama0407}, we transform a physical state given by the diagonalization of total Hamiltonian in the Nilsson basis into the spherical basis, in which one can perform more easily theoretical calculations. Then, {\it ab initio} Brueckner G-matrix calculated in the spherical basis is represented
in terms of  the deformed basis, which enables us to solve the deformed BCS in the following way.

With the expansion coefficient $B_{\alpha}$ from the deformed to the spherical basis, $
|\alpha \Omega_{\alpha}> ( = | N n_z \Lambda_{\alpha} \Omega_{\alpha} > ) =\sum_{a} B_{a}^{\alpha}~|a \Omega_{\alpha} >
$ with $a = ( n_r, l, j)$, the pairing potential $\Delta_{p}$ between a $\alpha$ and its time conjugate state ${\bar \alpha}$ is calculated as
\begin{eqnarray} \label{eq:gap}
&&\Delta_{\alpha p \bar{\alpha}p} = - {1 \over 2} {1 \over (2 j_a +1)^{1/2}}
\sum_{J, c }g_{{pair}}^{p} F_{\alpha a \alpha a}^{J0} F_{\gamma c \gamma c}^{J0}
 \nonumber \\
&& \times G(aacc,J)(2 j_c +1)^{1/2} (u_{1p_c}^* v_{1p_c} + u_{2p_c}^* v_{2p_c}) ~,
\end{eqnarray}
where $F_{ \alpha a  {\bar \beta b
}}^{JK'}=B_{a}^{\alpha}~B_{b}^{\beta} ~C^{JK'}_{j_{\alpha}
\Omega_{\alpha} j_{\beta}\Omega_{\beta}}$ is introduced for the transformation
to the deformed basis of G-matrix. Here $K'=\Omega_{\alpha}+\Omega_{\beta} $, which is a projection number
of the total angular momentum $J$ onto the $z$ axis, is selected $K'=0$ at the BCS stage
because we consider the pairings of the quasi-particles at $\alpha$ and ${\bar\alpha}$ states. In order to renormalize the G-matrix, strength parameters,
$g_{{pair}}^{p}$, $g_{{pair}}^{n}$ and $g_{{pair}}^{pn}$ are multiplied to the G-matrix \cite{Ha1}.
\begin{figure}
\includegraphics[width=1.0\linewidth]{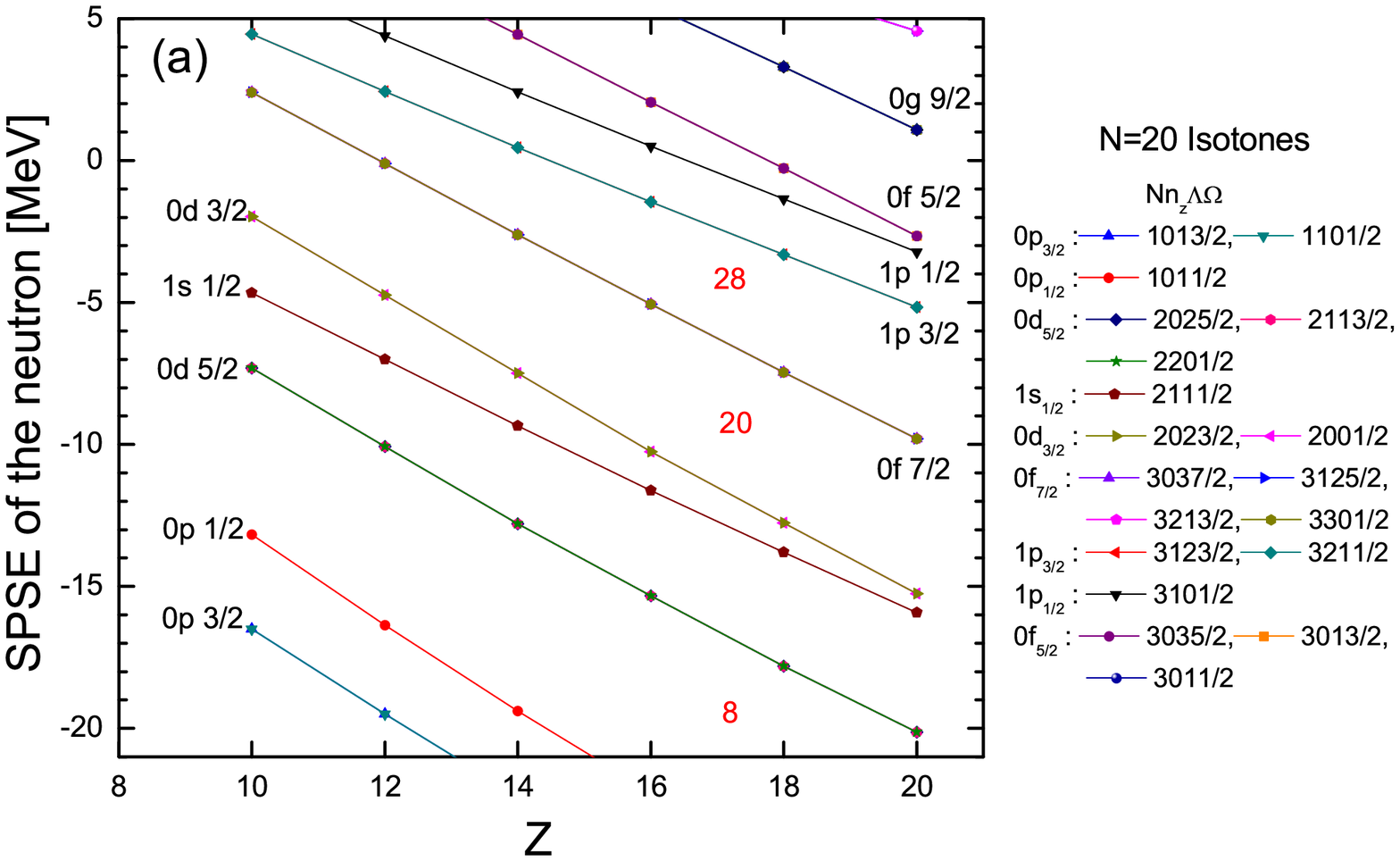}
\includegraphics[width=1.0\linewidth]{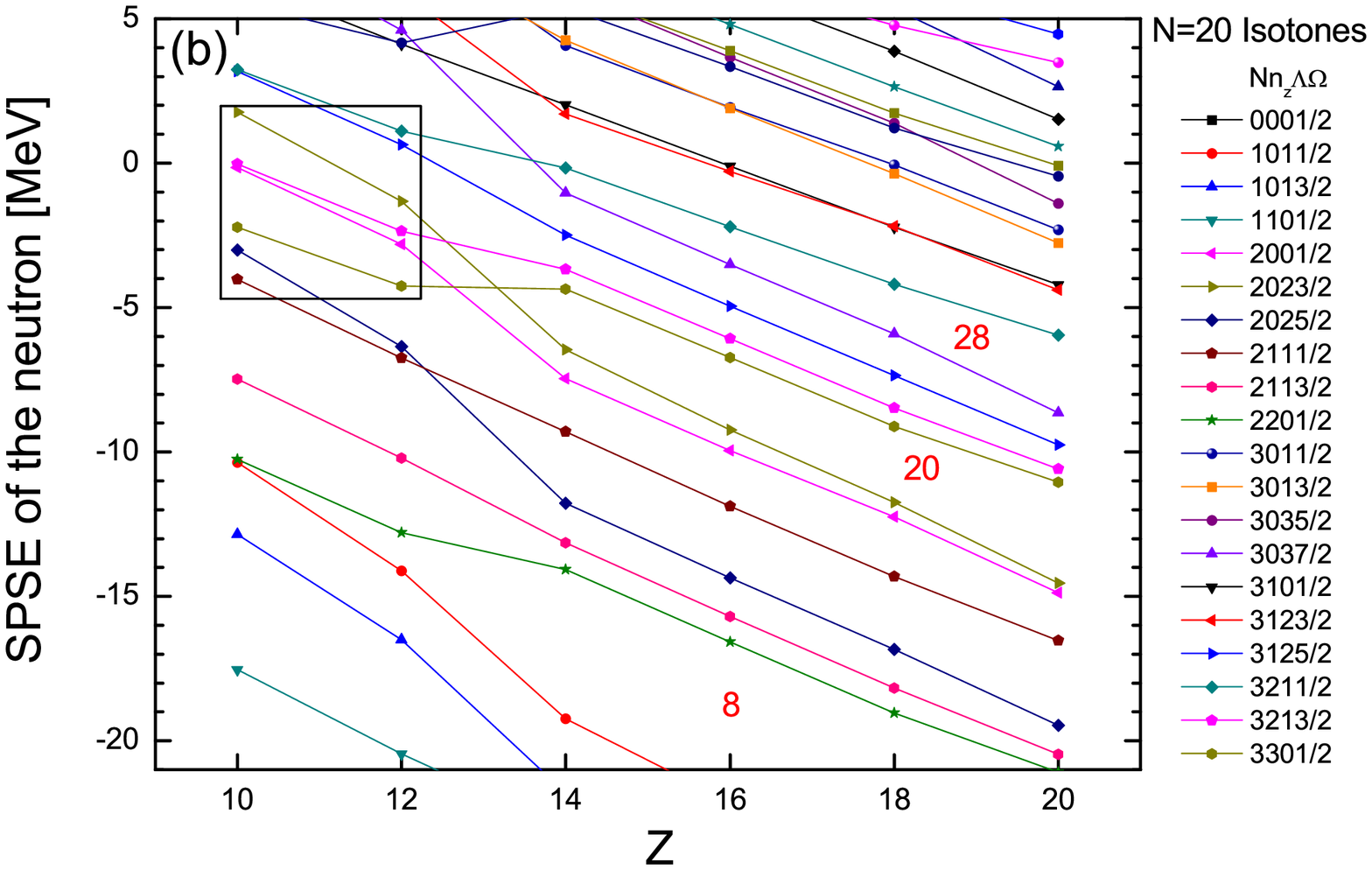}
\caption{(Color online) Neutron single particle state energy (SPSE) of N=20 isotones as a function of the proton number with the deformation parameter $\beta_2$
from the relativistic mean field (RMF) (upper panel) and from the B(E2) data \cite{Yana,Raman} (lower panel) tabulated in Table I. Since all $\beta_2^{RMF}$ values in the upper panel are zero, SPSEs by deformed basis are degenerated into spherical basis, whose degeneracy is explained in the legend of right hand side. Detailed evolutions from the degeneracy by the deformation are shown in the lower panel. The inside box in the lower panel stands for $f-$ intruder region.}
\label{fig1}
\end{figure}

\begin{figure}
\includegraphics[width=1.0\linewidth]{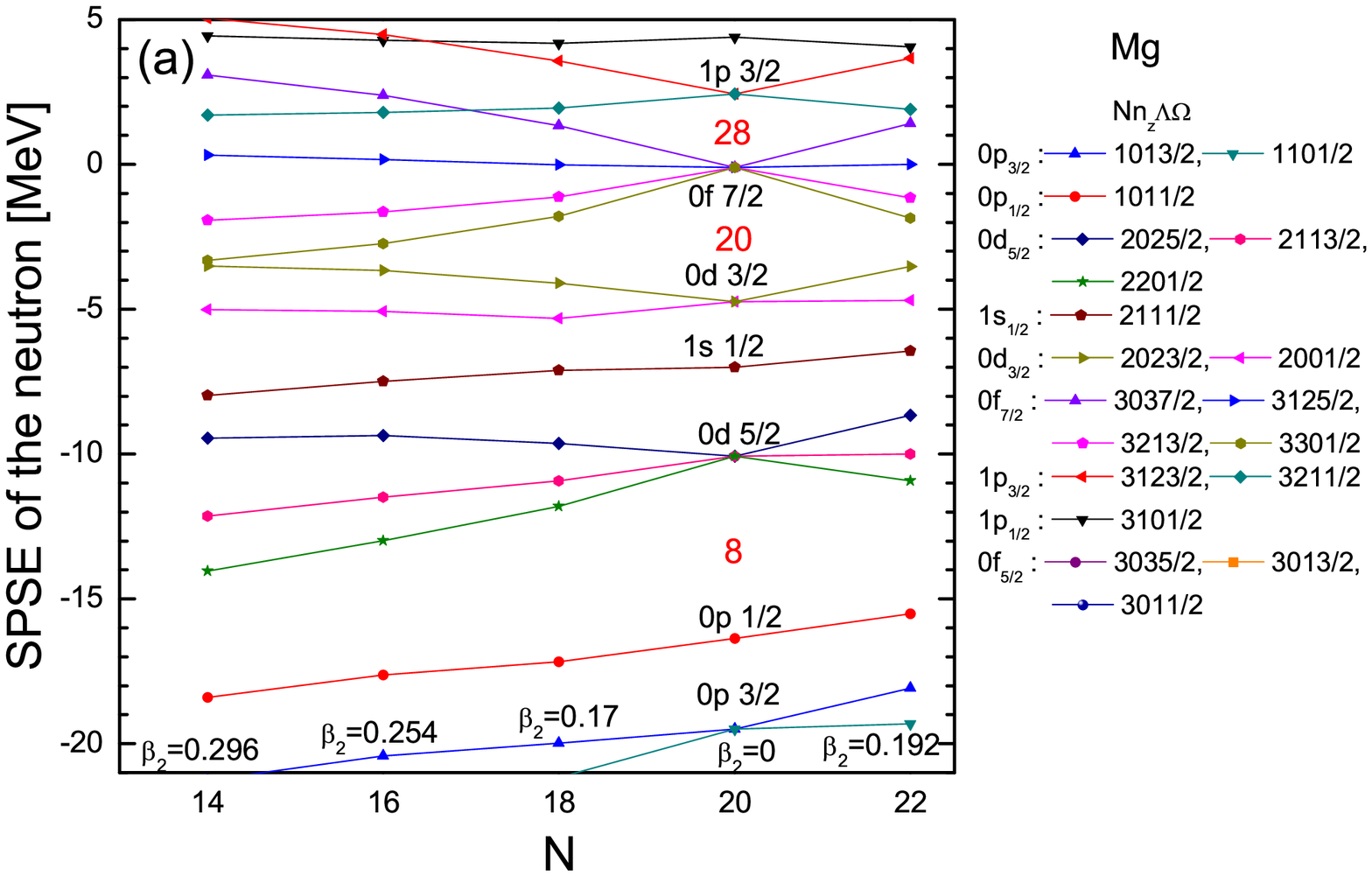}
\includegraphics[width=1.0\linewidth]{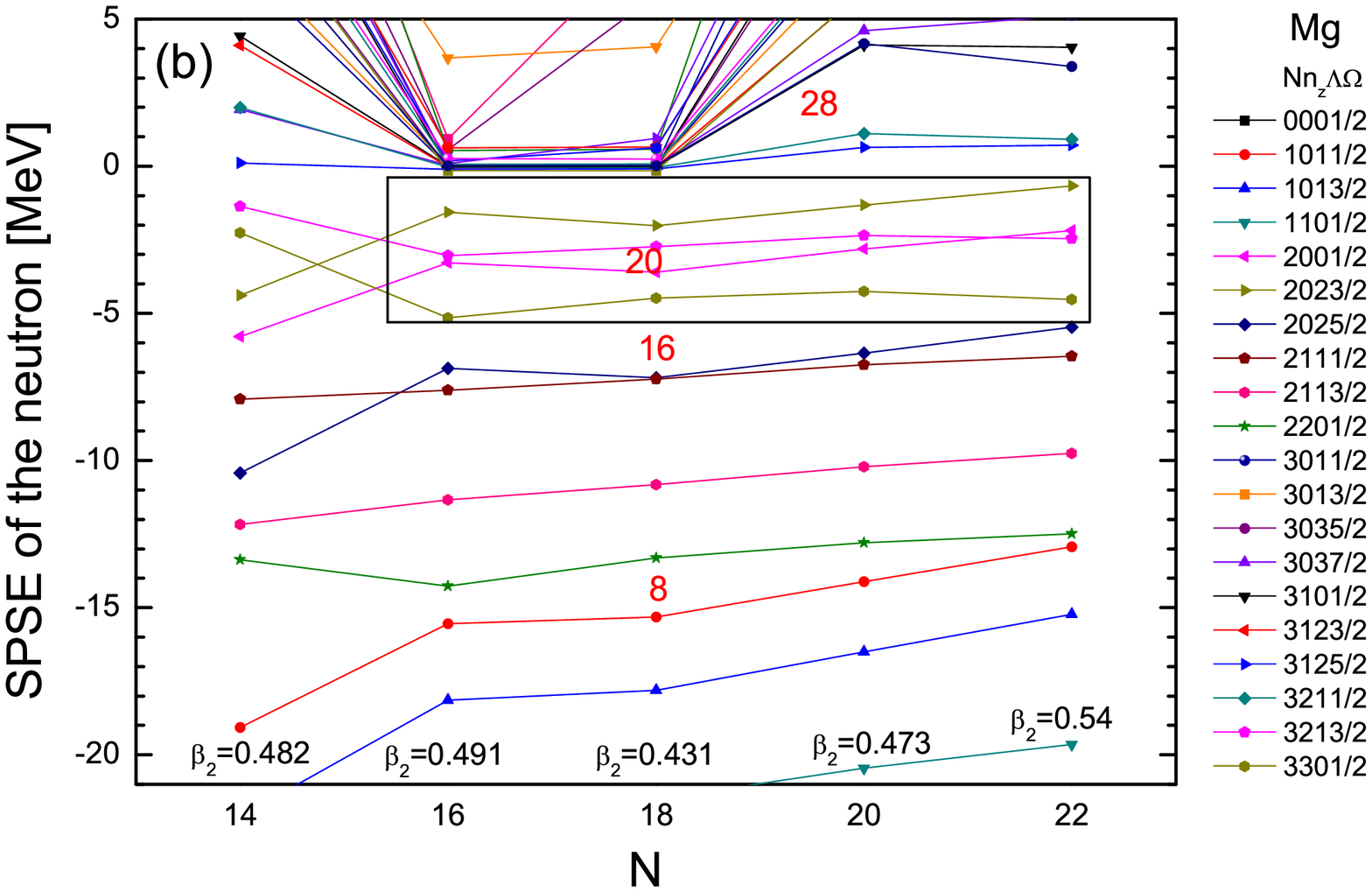}
\caption{(Color online) Same as Fig.1 but for Mg isotopes. The upper (lower) panel is for $\beta_2^{RMF} ( \beta_2^{E2})$ in Table II.}
\label{fig2}
\end{figure}

The $\beta^{\pm}$ transition amplitudes from the ground state of an initial nucleus
to the excited state are expressed by
\begin{eqnarray}
&&< 1(K),m | {\hat\beta}_{K }^- | ~QRPA >   \nonumber \\
&&= \sum_{\alpha \alpha''\rho_{\alpha} \beta \beta''\rho_{\beta}}{\cal N}_{\alpha \alpha''\rho_{\alpha}
 \beta \beta''\rho_{\beta} }
 < \alpha \alpha''p \rho_{\alpha}|  \sigma_K | \beta \beta''n \rho_{\beta}> \nonumber \\
&& \times [ u_{p \alpha \alpha''} v_{n \beta \beta''} X_{\alpha \alpha''  \beta \beta'',K} +
v_{p \alpha \alpha''} u_{n \beta \beta''} Y_{\alpha \alpha'' \beta \beta'',K}  ] ~,
\end{eqnarray}
where ${\cal N}$ is the nomalization factor and $|~QRPA >$ denotes the correlated QRPA ground state in the intrinsic frame.

To compare to the experimental data, the GT($\pm$) strength functions
\begin{equation}
B_{GT}^{\pm}(m)= \sum_{K=0,\pm 1} | < 1(K),m || {\hat \beta}_{K }^{\pm} || ~QRPA > |^2
\end{equation}
and their running sums $S_{GT}^{\pm} = \Sigma_m B_{GT}^{\pm}(m)$ are evaluated. The single particle states are used up to $4 \hbar \omega$ in the spherical limit.
Since the GT strength distributions turn out to be sensitive on the deformation parameter, $\beta_2$ \cite{Hama0407},
we took $\beta_2$ values from both the relativistic mean field (RMF)\cite{Lala99} and the E2 transition data, $\beta_2=(4\pi/3ZR_0^2)[B(E2\uparrow)/e^2]^{1/2}(R_0=1.2A^{1/3})$\cite{Yana,Raman,Church}.

In order to compare our single particle state energies (SPSEs) of N = 20 nuclei to those by other nuclear models, which show the shell evolution, we plotted SPSEs in Fig. 1, whose deformation parameters are tabulated in Table I.

The level density of {\it fp} shell at the upper panel in Fig.1 is increased with the
decrease of proton number similarly to the result of Ref. \cite{Yama04} which exploited HF with Skyrme interactions (see Fig. 2 at Ref. \cite{Yama04}) in the QRPA scheme to study the first $2^+$ states in $^{32}$Mg and $^{30}$Ne. This trend becomes manifest if we use the E2 deformation parameter (lower panel). Of course, the increased level density may lead to the anomaly of the B(E2) values as shown in Ref. \cite{Church}. In particular, N=20 shell gap is explicitly broken with the decrease of proton number by the intruder of $f_{7/2}$ state below $d_{3/2}$ state. Note the inside box in Fig. 1(b), where (330 1/2) state from $f_{7/2}$ state is located below (202 3/2) and (200 1/2) states from $d_{3/2}$ state at Z = 10 and 12 region. Recent calculation by the shell model (see Fig. 3 (a) at Ref. \cite{Otsuka10}) also shows such a trend, although the effect itself was smaller than ours.

\begin{table}
\caption[bb]{Deformation parameters $\beta_2$ of N=20 isotones from RMF \cite{Lala99} and B(E2)\cite{Yana,Raman} for N=20 isotones.
}
\setlength{\tabcolsep}{2.0 mm}
\begin{tabular}{ccc}\hline
           A & $\beta_2^{RMF}$ &   $\beta_2^{E2}$(B(E2))      \\ \hline \hline
 ${}^{30}$Ne &  0.  &  0.580       \\
 ${}^{32}$Mg &  0.  &  0.473       \\
 ${}^{34}$Si &  0.  &  0.179       \\
 ${}^{36}$S  &  0.  &  0.168       \\
 ${}^{38}$Ar &  0.  &  0.163       \\
 ${}^{40}$Ca &  0.  &  0.123       \\ \hline
 \end{tabular}
\label{tab:isotone}
\end{table}

Changes of SPSEs of Mg isotopes by the prolate deformation are presented in Fig. 2. In Table II, $\beta_2^{RMF}$, $\beta_2^{E2}$ and relevant pairing gaps are tabulated.
The empirical pairing gaps for protons and neutrons are evaluated by the odd-even mass differences through a symmetric five-term mass formula \cite{Audi}. In the lower panel, the (3301/2) state stemming from $f_{7/2}$ state is generated below Fermi surface energy, i.e. {\it f --} intruder states as shown at the inside box in Fig.2(b). As a result, we find the disappearance of the magic number N = 20 for N = 16 $\sim$ 22 nuclei. It is the most remarkable point in the lower panel, which used $\beta_2^{E2}$ accounting for the large B(E2) strength.
\begin{table}
\caption[bb]{Deformation parameters $\beta_2^{RMF}$ and $\beta_2^{E2}$, and empirical(theoretical) pairing gaps $\Delta^{{p,n}}_{{em}}$($\Delta^{{p,n}}_{{th}}$) for Mg isotopes. The deformation parameter $\beta_2^{E2} $ for $ ^{26}$Mg $\sim ^{32}$Mg and $^{34}$Mg are from the B(E2) data in Refs. \cite{Raman} and \cite{Church}, respectively.
The particle-particle (particle-hole) strength parameters are taken as $g_{{pp}}=1.0 (g_{{ph}}=1.0)$.
}
\setlength{\tabcolsep}{2.0 mm}
\begin{tabular}{cccccc}\hline
 A & $\beta_2^{RMF}$ & $\beta_2^{E2} $  & $\Delta^{p}_{em}$ ($\Delta^{p}_{th})$& $\Delta^{n}_{em}$($\Delta^{n}_{th}$) & $\Delta^{np}_{em}$
  \\ \hline \hline
 ${}^{30}$Mg &  0.17   &  0.431      & 2.094(2.095)  & 1.719(1.723) & 0.61   \\
 ${}^{32}$Mg &  0.1    &  0.473      & 2.510(2.522)   & 1.560(1.576) & 0.64  \\
 ${}^{34}$Mg &  0.192  &  0.54       & 2.980(2.991)   & 1.330(1.331) & 0.89  \\ \hline
 \end{tabular}
\label{tab:isotope}
\end{table}
These shell evolution phenomena by the intruder are intimately related to the following facts. The SPSEs adopted from the deformed Woods Saxon potential naturally depend on the parameter $\beta_2$. The deformation of nuclei may be conjectured to come from macroscopic phenomena, for example, the core polarization, the high spin states and so on. Microscopic reasons may be traced to the tensor force in nucleon-nucleon interaction, which is known to account for the shell evolution according to the recent shell model calculations \cite{Otsuka05,Otsuka10}.  For example, T = 0, J = 1 pairing, which is associated with the $^{3}S_0$ tensor force, may lead to the deformation compared to the T = 1, J = 0 pairing. Therefore, the deformation parameter adopted in this work may include implicitly and effectively such effects, because the single particle states from the deformed Wood Saxon potential show strong dependences on the $\beta_2$, as shown in Figs. 1 and 2.

Now we discuss the GT strengths on Mg isotopes whose SPSEs heavily depend on the neutron numbers as shown in Fig. 2. In particular, $^{32}$Mg is thought to locate in the island inversion. Since the nuclei considered here are expected to have large energy gaps between proton and neutron spaces, similarly to the neutron-rich nuclei of importance in the r-process, we consider only the $nn$ and $pp$ pairing correlation, which have only isospin T = 1 and J = 0 interaction.

\begin{figure}
\includegraphics[width=1.0\linewidth]{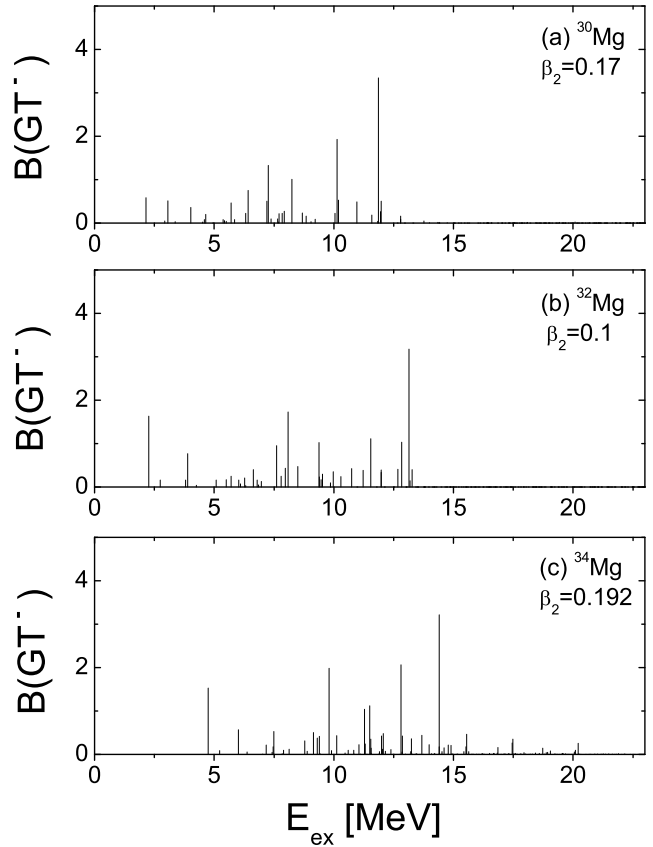}
\caption{(Color online) Gamow-Teller strength distributions B(GT$^{-}$) on Mg isotopes as a function of the excitation
energy $E_{ex}$ w.r.t. the ground state of $^{30}$Mg $\sim ^{34}$Mg with the deformation parameter $\beta_2^{RMF}$ at the upper panel in Fig. 2.}
\label{fig3}
\end{figure}
\begin{figure}
\includegraphics[width=1.0\linewidth]{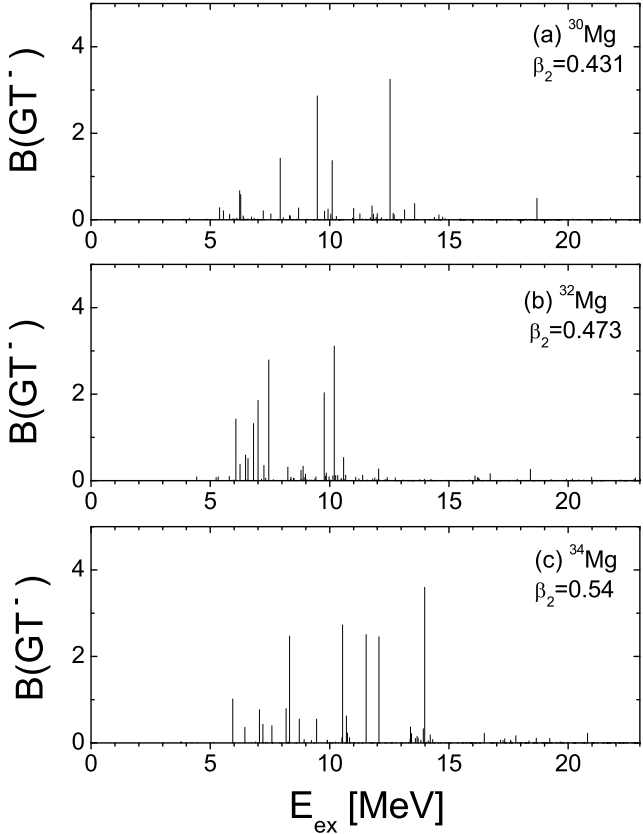}
\caption{(Color online) Same as Fig. 3 but with the deformation parameter $\beta_2^{E2}$ used at the lower panel in Fig. 2.}
\label{fig4}
\end{figure}

In Fig. 3, the GT strength distributions by $\beta_2^{RMF}$, B(GT$^-$), on Mg isotopes are presented as a
function of the excitation energy $E_{ex}$ w.r.t. the ground state of $^{30}$Mg $\sim ^{34}$Mg for $\beta_2^{RMF}$ values \cite{Lala99}.
The GT strength distributions are widely scattered into the HL states owing to the deformation.
Actually, these HL GT states are intimately associated with the 2p-2h contributions which result from the wide smearing of some physical states around the Fermi surface by the prolate deformation \cite{Ha1,Ha2}.

Fig.4 is the results by $\beta_2^{E2}$, which shows an interesting point. Some low-lying GT states below 5 MeV disappeared if we use $\beta_2^{E2}$ values.
%
%
\begin{figure}
\includegraphics[width=1.0\linewidth]{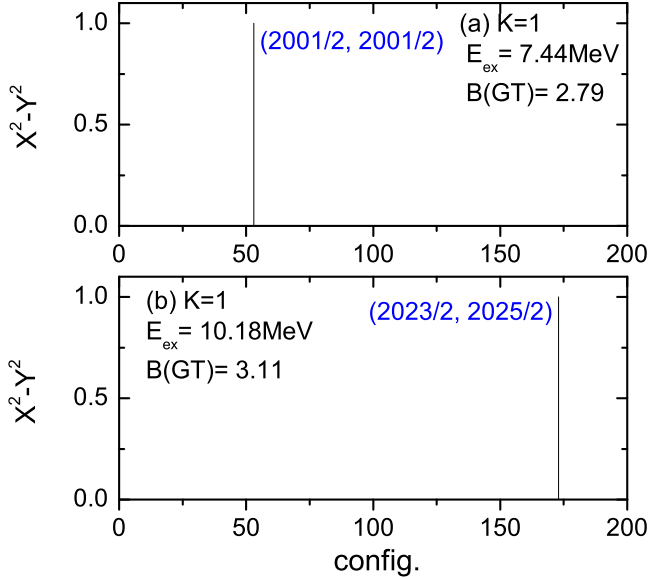}
\caption{(Color online) Main collective states presented by $X^{2}-Y^{2}$ for two dominant GT states w.r.t. the ground state of $^{32}$Mg as a function of the configuration state for the case of Fig.4 (b).}
\label{fig5}
\end{figure}
%
%
%
This can be explained if we recollect the disappearance of the magic number $N = 20$ in Fig.2(b). For detailed analysis, in Fig.5, we identify main collective $1^+$ states for the GT transitions on $^{32}$Mg for $\beta_2^{E2} =0.473$, which are presented as $X^{2}-Y^{2}$ of two excitation energies having relatively larger B(GT) values in Fig.4 (b). $X(Y)$ is a forward(backward) amplitude in Eq. (2).
The transition at 7.44 MeV in K = 1 come from the combination of two (2001/2) states and the transition at 10.18 MeV turns out to come from the combination of (2203/2) and (2025/2) sates.
An important point to be noticed in Fig.5 is that the (2001/2) state stemming from the $d_{3/2}$ state may contribute to the GT transition at 7.44 MeV by the $f-$ intruder state.
These HL GT states, which were usually not considered because of the limited models space in theoretical side and the limited incident energy of projectiles in experimental side, enable us to abandon the quenching factor for the Ikeda sum rule and to take into account of HL GT states in the neutrino-induced reaction, in particular, by the supernova neutrino, whose energy range may extend up to tens of MeV \cite{Ch13}.

In summary, to describe single particle states in deformed basis, we exploited the deformed axially symmetric Woods-Saxon potential, and performed the deformed BCS and deformed QRPA with the realistic two-body interaction recalculated in deformed basis from {\it ab initio} G-matrix based on Bonn CD potential. Results of the Gamow-Teller strength for Mg isotopes show that the deformation effect leads to a fragmentation of the GT strength into HL states and predicts the HL GT excitations to be explored by higher energy projectiles, whose excitations may affect the neutrino process in the cosmos through the GT transition. They are shown to result from the wide smearing by the increase of the Fermi surface energy due to the prolate deformation.

In order to properly describe the deformed nuclei, in particular, located near to stability region, the T= 0 and J = 1 {\it np} pairing should be also taken into account
because the J = 1 pairing is believed as one of the reasons leading to the nuclear deformation. Also, for the p-process nuclei, the {\it np} pairing could be more important than the neutron-rich nuclei because of the adjacent
energy gaps of protons and neutrons. The calculations for the neutron-deficient nuclei in the p-process are in progress. Since the continuum states are of importance in neutron-rich nuclei, we are also preparing to include the continuum states in the present deformed QRPA.

This work was supported by the National Research Foundation of Korea (2012R1A1A3009733, 2012-009733 and 2011-0015467).

\end{document}